\documentclass[10pt,final, twocolumn]{IEEEtran}
\usepackage{graphicx,cite,epsfig,amssymb,amsmath}
\usepackage[font=small,labelsep=none, justification=centering]{caption}
\usepackage{algorithm} 
\usepackage{algorithmic}
\usepackage{amsmath}
\usepackage{xcolor}

\begin{document}
\title{\mbox{}\vspace{0.40cm}\\
\textsc{Enabling security and High Energy Efficiency in the Internet of Things with Massive MIMO Hybrid Precoding} \vspace{0.2cm}}

\author
{
    \IEEEauthorblockN
    {
        Ran Zi\IEEEauthorrefmark{1}\IEEEauthorrefmark{2},
        Jia Liu\IEEEauthorrefmark{1},
        Liang Gu\IEEEauthorrefmark{2} and
        Xiaohu Ge\IEEEauthorrefmark{3}
    }


    \IEEEauthorblockA
    {
        \IEEEauthorrefmark{1}Shenzhen Institutes of Advanced Technology, \\
        Chinese Academy of Sciences, Shenzhen, China\\
    }

    \IEEEauthorblockA
    {
        \IEEEauthorrefmark{2}Sangfor Technologies, Shenzhen, China\\
    }

    \IEEEauthorblockA
    {
        \IEEEauthorrefmark{3}School of Electronics \& Information Engineering\\
        Huazhong University of Science \& Technology, Wuhan, China\\
    }

}


\maketitle
\renewcommand{\thefootnote}{}

\footnote{This work was supported by
Shenzhen Science and Technology Innovation Commission (Grant No. JCYJ20160608153506088).}

\begin{abstract}
Recently, the security of Internet of Things (IoT) has been an issue of great concern. Physical layer security methods can help IoT networks achieve information-theoretical secrecy. Nevertheless, utilizing physical security methods, such as artificial noise (AN) may cost extra power, which leads to low secure energy efficiency. In this paper, the hybrid precoding technique is employed to improve the secure energy efficiency of the IoT network. A secure energy efficiency optimization problem is formulated for the IoT network. Due to the non-convexity of the problem and the feasible domain, the problem is firstly transformed into a tractable suboptimal form. Then a secure hybrid precoding energy efficient (SEEHP) algorithm is proposed to tackle the problem. Numerical results indicate that the proposed SEEHP algorithm achieves higher secure energy efficiency compared with three existing physical layer security algorithms, especially when the number of transmit antennas is large.

\end{abstract}
\vspace{2 ex}

\begin{IEEEkeywords}
Internet of Things, physical layer security, hybrid precoding, secure energy efficiency, secrecy capacity
\end{IEEEkeywords}


%
\IEEEpeerreviewmaketitle

\section{Introduction}
\label{sec1}
The IoT has drawn great interests in both academic and industrial areas. IoT is supposed to enable ubiquitous connectivity among billions of physical objects, such as vehicles, mobile phones, sensors, etc \cite{Gubbi13,Fuqaha15}. IoT greatly promotes the gathering and exchanging of information, which is believed to bring a huge boost in productivity and act as a cornerstone in the future intelligent society. In recent years, the large-scale application of IoT is achieved with the rapid development of wireless communication technology, cloud computing, integrated circuits, etc \cite{Lin17}.

Since IoT is widely adopted in transportation, medical health, industrial and even military fields, the security of IoT is greatly concerned \cite{Weber10,Keoh14}. Specifically, messages baring high values are transmitted among large amounts of IoT devices. Any leakage of these sensitive messages due to eavesdropping may lead to unacceptable consequences. A widely used method to prevent information eavesdropping is cryptographic encryption \cite{Keoh14,Granjal15,Zhang15,Mukherjee15}. However, the cryptographic encryption method is based on the distribution and management of the secret keys, which is too complex and power-consuming in IoT networks \cite{Granjal15,Zhang15}. Because IoT networks consist of large amounts of IoT devices with constrained resources (e.g. energy, storage, computing), the easy adoption and energy efficiency of the secure transmission method are key requirements \cite{Mukherjee15}. Additionally, with the further adoption of quantum computing and other high performance computing technologies, the secrecy level of the cryptographic encryption method is also reduced, since the traditional cryptographic method can be easily cracked with enough computing capability \cite{Bernstein09,Stinson05}. In this case, secure transmission methods, which can be easily implemented in IoT networks and prevent the message from ciphertext cracking, are urgently needed.

Different from traditional cryptographic methods, in recent years, physical layer security has been proposed as a key-less secure transmission method whose secrecy is guaranteed by information theory \cite{Mukherjee15,Wyner75,Csisza78,Shafiee09}. Exploiting the inherent randomness of the physical transmission medium and the difference between the legitimate channel and the wiretap channel, information-theoretical secrecy is perfectly achieved when the quality of the wiretap channel is lower than the desired legitimate channel \cite{Wyner75,Csisza78}. In the pioneering works by Wyner, the wiretap channel model and the secrecy capacity are introduced when the eavesdropper's channel is a degraded version of the legitimate receiver's channel \cite{Wyner75}. The secrecy capacity is the largest rate communicated between the source and legitimate receiver with the eavesdropper knowing no information of the messages. Then, Csiszar and Korner considered a more general scenario and studied the secrecy capacity where the channel is not degraded \cite{Csisza78}. Other authors extended secrecy capacity investigation to Gaussian MIMO channels \cite{Shafiee09,Yang13}, and studied the impact of fading channels on the secrecy capacity \cite{Gopala08}. With physical layer security, secure transmission is guaranteed without encryption or decryption computation at the IoT devices, which eliminates the risk of cracking and makes IoT networks more energy efficient.

Among the existing researches on physical layer security, utilizing multiple-input multiple-output (MIMO) and artificial noise (AN) technologies to improve the legitimate channel capacity and degrade the wiretap channel has been studied \cite{Chen15_16,Zhu14_17,Chen14_18,Goel08_19,Wang15_20,HMWang15_21,HMWang16_22}. The authors in \cite{Chen15_16} investigate the secure limiting performance of the massive MIMO system when the number of antennas approaches infinity. For multi-user scenarios, secure massive MIMO transmission with imperfect channel state information (CSI) has been investigated \cite{Zhu14_17}. In \cite{Chen14_18}, the secrecy outage performance for massive MIMO relaying systems are studied with imperfect channel state information. The authors in \cite{Goel08_19} uses AN to degrade the eavesdropper's channel when the transmitter and the relay are equipped with multiple antennas. AN assisted secure massive MIMO in Rician channels is studied in \cite{Wang15_20}, where the secrecy outage is defined and derived to study the impact of the eavesdropper's location. In \cite{HMWang15_21}, a thorough investigation and optimization is proposed for AN assisted secure MIMO systems. Combining AN and multi-cell multi-user MIMO systems under a stochastic geometry framework, a comprehensive performance analysis is provided in \cite{HMWang16_22}.

Nevertheless, transmitting AN and utilizing secure precoding schemes may cost extra power. How to increase secrecy capacity while maintaining low power consuming has drawn people's attention \cite{DWang16_23,HZhang14_24,DWang16_25,Zappone16_26}. In this case, the secure energy efficiency is defined as the secret bits transferred with unit energy \cite{DWang16_23}. The secure energy efficiency is modelled as a nonconvex optimization function and maximized by joint source and relay power allocation in \cite{DWang16_23}. In \cite{HZhang14_24}, the authors maximize the energy efficiency of the three-node MIMO system with an eavesdropper subject to the secret rate and transmit power constraints. An iterative algorithm with is proposed to improve the secure energy efficiency. In \cite{DWang16_25}, the secure energy efficiency for the untrusted two-way relaying network is investigated and optimized. The nonconvex optimization problem is tackled and the proposed algorithm remarkably increases the secure energy efficiency yet at the cost of secrecy sum rate loss. Utilizing AN and MIMO to increase the secrecy capacity, energy-efficient resource allocation in multiple-antenna wiretap channels is investigated in \cite{Zappone16_26}. The results show that AN does not always improve the system secure energy efficiency, depending on the digital signal processor used to compute the resource allocation.

As introduced above, various transmission strategies are designed and investigated to improve the secrecy capacity and energy efficiency with MIMO and AN. Meanwhile, hybrid precoding has been recently proposed for MIMO systems because of its low hardware complexity and high energy efficiency \cite{Alkhateeb14,Ge18,Liu14,Zi16}. In hybrid precoding schemes, the number of radio frequency (RF) chains is less than the number of antennas, which leads to tremendous hardware complexity reduction and energy efficiency improvement, especially for massive MIMO systems with a huge number of antennas. In \cite{Alkhateeb14}, based on a low-complexity channel estimation algorithm, a hybrid precoding algorithm was proposed to achieve sub-optimal performance in the single-user system. The authors in \cite{Ge18} investigated a joint optimization problem of computation and communication power for massive MIMO systems with partially-connected RF chains. Considering the limited number of RF chains and the phase-only constraint, a two stage precoding scheme was proposed to exploit channel gains provided by the spatial degrees of freedom in massive MIMO systems \cite{Liu14}. To reduce the costs of RF chains at the base station and improve the energy efficiency, the authors in \cite{Zi16} firstly proposes an energy efficient hybrid precoding scheme for massive MIMO systems.

However, to the best of our knowledge, it is still an open issue to implement hybrid precoding MIMO in IoT networks to improve the secure energy efficiency. So in this paper, considering the gateway controller of the IoT network equipped with a hybrid precoding massive MIMO antenna array, the secure energy efficiency optimization problem is investigated in order to obtain both high secrecy capacity and low power consumption. The contributions of this paper are as follows.

\begin{enumerate}
\item The secure energy efficiency optimization problem of the gateway controller in the IoT network is formulated, considering the transmit power of the information bearing signal and the AN, as well as the power consumed by the gateway controller hardware, such as the RF chains, phase shifters and constant power consumption.
\item To tackle the non-convexity of the objective function and the feasible domain, an RF and baseband precoding scheme is designed and implemented to transform the optimization problem into a more trackable suboptimal form.
\item A secure energy efficiency hybrid precoding algorithm is proposed to solve the transformed optimization problem. Numerical results indicate that the proposed SEEHP algorithm achieves the highest secure energy efficiency and close secrecy capacity compared with other three physical layer security algorithms when the maximum transmit power is small. In addition, when the number of antennas keeps increasing, the secure energy efficiency performance advantage of the proposed SEEHP algorithm strengthens.
\end{enumerate}

The remainder of this paper is outlined as follows. In Section II, the system model is introduced. The secure energy efficiency optimization problem is formulated in Section III. Section IV detailedly describes how the secure energy efficiency optimization problem is transformed and solved with the proposed SEEHP algorithm. Section V gives the numerical results of the SEEHP algorithm compared with other three physical layer security algorithms. Conclusions are drawn in Section VI.

{\em Notations:} Bold upper (or lower) letters appear in the following sections denote matrices (or column vectors). ${\bf{X}} \in {\mathbb{C}^{A \times B}}$ denotes that ${\bf{X}}$ is a matrix with $A$ rows and $B$ columns. ${{\left( \centerdot  \right)}^{H}}$ denotes the conjugate transpose. $\left| \left( \centerdot  \right) \right|$ denotes the absolute value. $\left\| \left( \centerdot  \right) \right\|$ denotes the L-2 norm of a vector. $\sim $ is used to denote the equality in a distribution. $\overset{\lower0.5em\hbox{$\smash{\scriptscriptstyle\frown}$}}{X}$ and $\overset{\lower0.5em\hbox{$\smash{\scriptscriptstyle\smile}$}}{X}$ denote the upper and lower bound of the variable $X$, respectively. ${{\left( \centerdot  \right)}^{+}}$ is used to denote $\max \left\{ 0,\left( \centerdot  \right) \right\}$.

\section{SYSTEM MODEL}
\label{sec2}

\begin{figure}
  \centering
  \includegraphics[width=9.2cm,draft=false]{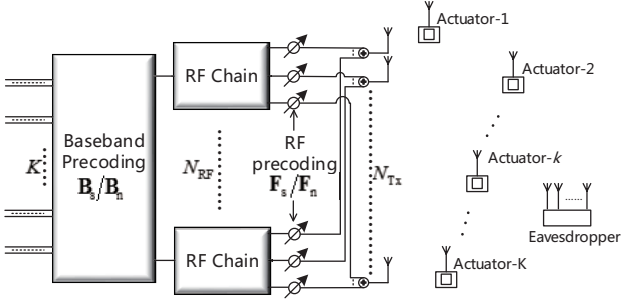}\\
  \caption{\small \ \ System model.}\label{fig1}
\end{figure}

In this paper, we consider an IoT network in which a gateway controller sends confidential messages to $K$ single-antenna actuators, as depicted in Fig. 1. The gateway controller is equipped with a massive MIMO antenna array with ${N_{{\text{Tx}}}}$ antennas, ${N_{{\text{Tx}}}} \gg K$. To reduce the hardware complexity and energy consumption of the IoT network, the gateway controller antenna array is assumed to be connected to ${N_{{\text{RF}}}}$ radio frequency chains and simultaneously transmits $K$ information baring data streams, $K < {N_{{\text{RF}}}} < {N_{{\text{TX}}}}$. Meanwhile, there exists an eavesdropper attempting to overhear the signal transmitted from the gateway controller R to the $k{\text{th}}$ actuator. Assuming the eavesdropper is equipped with $M$ antennas to enhance the eavesdropping capability, $M > {N_{{\text{RF}}}} - K$.

Denote ${\mathbf{s}} \in {\mathbb{C}^{K \times 1}}$ as the baseband symbol vector transmitted from the gateway controller to $K$ actuators. In order to degrade the overheard signal at the eavesdropper, the gateway controller emits the AN ${\mathbf{n}} \in {\mathbb{C}^{{N_{\text{E}}} \times 1}}$ using the ${N_{\text{E}}}{\text{ = }}{N_{{\text{RF}}}} - K$ remaining degrees of freedom offered by the ${N_{{\text{RF}}}}$ RF chains. Without loss of generality, assume that the entries of ${\mathbf{s}}$ and ${\mathbf{n}}$ are independent and identically distributed (i.i.d.) Gaussian random variables with zero mean and variance of 1. The information baring signals and the AN are firstly processed by the baseband digital precoders ${{\mathbf{B}}_{\text{s}}} \in {\mathbb{C}^{{N_{{\text{RF}}}} \times K}}$ and ${{\mathbf{B}}_{\text{n}}} \in {\mathbb{C}^{{N_{{\text{RF}}}} \times {N_{\text{E}}}}}$, respectively. Afterwards, they are up-converted to the RF domain by the ${N_{{\text{RF}}}}$ RF chains and processed by the RF analog precoders ${{\mathbf{F}}_{\text{s}}} \in {\mathbb{C}^{{N_{{\text{Tx}}}} \times {N_{{\text{RF}}}}}}$ and ${{\mathbf{F}}_{\text{n}}} \in {\mathbb{C}^{{N_{{\text{Tx}}}} \times {N_{{\text{RF}}}}}}$ via phase shifters.

The downlink channels between the gateway controller and the $K$ actuators are assumed to be i.i.d. quasi-static flat Rayleigh channels \cite{Chen15_16,Zhu14_17,Chen14_18}. In this case, the channel only involves the none-line-of-sight (NLoS) propagation without any line-of-sight (LOS) components \cite{Chen14_18}. Further denote the downlink channel between the gateway controller and the $k{\text{th}}$ actuator as ${\mathbf{h}}_k^H = \sqrt {{\beta _k}} {\mathbf{g}}_k^H \in {\mathbb{C}^{1 \times {N_{{\text{Tx}}}}}}$, where ${\beta _k}$ and ${\mathbf{g}}_k^H$ represent the path-loss and small-scale fading component between the gateway controller and the actuator, respectively. ${\beta _k} = {\kappa  \mathord{\left/
 {\vphantom {\kappa  {l_k^\chi }}} \right.
 \kern-\nulldelimiterspace} {l_k^\chi }}$ , where $\kappa $ is the lognormal random variable with the zero mean and the variance 9.2 dB, ${l_k}$ is the distance between the BS and the kth actuator, and $\chi $ is the path loss exponent. The downlink channel between the gateway controller and the eavesdropper is similarly denoted as ${\mathbf{H}}_{\text{E}}^H = \sqrt {{\beta _{\text{E}}}} {\mathbf{G}}_{\text{E}}^H \in {\mathbb{C}^{M \times {N_{{\text{Tx}}}}}}$, where ${\beta _{\text{E}}}$ and ${\mathbf{G}}_{\text{E}}^H$ represent the path-loss and small-scale fading component between the gateway controller and the eavesdropper, respectively. ${\beta _{\text{E}}} = {\kappa  \mathord{\left/
 {\vphantom {\kappa  {l_E^\chi }}} \right.
 \kern-\nulldelimiterspace} {l_E^\chi }}$, where ${l_E}$ is the distance between the BS and the kth actuator. Based on the assumption that the downlink channels are i.i.d. quasi-static flat Rayleigh channels, the entries of the small-scale fading vector ${\mathbf{g}}_k^H$ and the small-scale fading matrix ${\mathbf{G}}_{\text{E}}^H$ are all i.i.d. complex Gaussian variables. Thus, the received signal at the $k{\text{th}}$ actuator is written as

\[{y_k} = {\mathbf{h}}_k^H{{\mathbf{F}}_{\text{s}}}{{\mathbf{B}}_{\text{s}}}{\mathbf{s}} + {\mathbf{h}}_k^H{{\mathbf{F}}_{\text{n}}}{{\mathbf{B}}_{\text{n}}}{\mathbf{n}} + {w_k},\tag{1}\]
where ${w_k}$ is the additive white complex Gaussian noise with zero-mean and variance $\sigma _w^2$.

The received signal at the eavesdropper is written as

\[{{\mathbf{y}}_{\text{E}}} = {\mathbf{H}}_{\text{E}}^H{{\mathbf{F}}_{\text{s}}}{{\mathbf{B}}_{\text{s}}}{\mathbf{s}} + {\mathbf{H}}_{\text{E}}^H{{\mathbf{F}}_{\text{n}}}{{\mathbf{B}}_{\text{n}}}{\mathbf{n}} + {{\mathbf{w}}_{\text{E}}},\tag{2}\]
where ${{\mathbf{w}}_{\text{E}}} \in {\mathbb{C}^{M \times 1}}$ is the additive white complex Gaussian noise vector whose entries follows i.i.d. complex Gaussian distributions with zero-mean and variance $\sigma _w^2$. In this case, the downlink signal power to the $k{\text{th}}$ actuator is ${P_k} = {\left\| {{{\mathbf{F}}_{\text{s}}}{{\mathbf{b}}_{{\text{s,}}k}}{s_k}} \right\|^2} = {\left\| {{{\mathbf{F}}_{\text{s}}}{{\mathbf{b}}_{{\text{s,}}k}}} \right\|^2}$ and the total signal power is $\sum\limits_{k = 1}^K {{{\left\| {{{\mathbf{F}}_{\text{s}}}{{\mathbf{b}}_{{\text{s,}}k}}} \right\|}^2}} $, where ${{\mathbf{b}}_{{\text{s,}}k}}$ is the $k{\text{th}}$ column of ${{\mathbf{B}}_{\text{s}}}$. The power of the emitted AN is ${P_{\text{n}}} = {\left\| {{{\mathbf{F}}_{\text{n}}}{{\mathbf{B}}_{\text{n}}}{\mathbf{n}}} \right\|^2} = {\left\| {{{\mathbf{F}}_{\text{n}}}{{\mathbf{B}}_{\text{n}}}} \right\|^2}$. Without loss of generality and similar to \cite{Jeong12,Cumanan14,Zheng15}, the gateway controller is assumed to have perfect knowledge of all the channels to the actuators and the eavesdropper. The assumption is practical when the eavesdropper is active in the IoT network, which is very realistic if the eavesdropper plays double roles, i.e. at one time as the legitimate nodes and at another time as eavesdroppers. The authors have also noted that there exist recent studies on imperfect channel knowledge in physical layer secure transmission. In this case, we focus on the scenario with perfect channel knowledge at the gateway controller. The imperfect channel knowledge scenario will be investigated in our future work.

\section{Problem formulation}
\label{sec3}
Based on the pre-described system model, the received SINR at the $k{\text{th}}$ actuator is

\[{\gamma _k} = \frac{{{\mathbf{h}}_k^H{{\mathbf{F}}_{\text{s}}}{{\mathbf{b}}_{{\text{s,}}k}}{\mathbf{b}}_{{\text{s}},k}^H{\mathbf{F}}_{\text{s}}^H{{\mathbf{h}}_k}}}{{\sum\limits_{i = 1,i \ne k}^K {{\mathbf{h}}_k^H{{\mathbf{F}}_{\text{s}}}{{\mathbf{b}}_{{\text{s,}}i}}{\mathbf{b}}_{{\text{s}},i}^H{\mathbf{F}}_{\text{s}}^H{{\mathbf{h}}_k}}  + {\mathbf{h}}_k^H{{\mathbf{F}}_{\text{n}}}{{\mathbf{B}}_{\text{n}}}{\mathbf{B}}_{\text{n}}^H{\mathbf{F}}_{\text{n}}^H{{\mathbf{h}}_k} + \sigma _w^2}}.\tag{3}\]

The capacity of the $k{\text{th}}$ actuator is then obtained by ${C_k} = {\log _2}\left( {1 + {\gamma _k}} \right)$. Moreover, we assume the worst case for secure IoT network that the eavesdropper is noiseless \cite{Zhou10}. In this case the received SINR at the eavesdropper achieves the upper bound \cite{Zhou10}

\[{\overset{\lower0.5em\hbox{$\smash{\scriptscriptstyle\frown}$}}{\gamma } _{{\text{E}},k}} = {\mathbf{b}}_{{\text{s,}}k}^H{\mathbf{F}}_{\text{s}}^H{{\mathbf{H}}_{\text{E}}}{\left( {{\mathbf{H}}_{\text{E}}^H{{\mathbf{F}}_{\text{n}}}{{\mathbf{B}}_{\text{n}}}{\mathbf{B}}_{\text{n}}^H{\mathbf{F}}_{\text{n}}^H{{\mathbf{H}}_{\text{E}}}} \right)^{ - 1}}{\mathbf{H}}_{\text{E}}^H{{\mathbf{F}}_{\text{s}}}{{\mathbf{b}}_{{\text{s,}}k}}.\tag{4}\]

The upper bound of the capacity of the eavesdropper is then obtained as ${\overset{\lower0.5em\hbox{$\smash{\scriptscriptstyle\frown}$}}{C} _{{\text{E}},k}} = {\log _2}\left( {1 + {{\overset{\lower0.5em\hbox{$\smash{\scriptscriptstyle\frown}$}}{\gamma } }_{{\text{E}},k}}} \right)$. Furthermore, the lower bound of the secrecy capacity of the $k{\text{th}}$ AN is derived as \cite{Yang13,Zhu14_17}

\[\overset{\lower0.5em\hbox{$\smash{\scriptscriptstyle\smile}$}}{C} _k^{{\text{sec}}} = {\left[ {{C_k} - {{\overset{\lower0.5em\hbox{$\smash{\scriptscriptstyle\frown}$}}{C} }_{{\text{E}},k}}} \right]^ + },\tag{5}\]
where ${\left[ x \right]^ + } = \max \left\{ {0,x} \right\}$. Because the eavesdropper only overhears the $k{\text{th}}$ actuator, for other actuators the secrecy capacity equals to the capacity, i.e. $C_l^{\sec } = {C_l} = {\log _2}\left( {1 + {\gamma _l}} \right)$, $l \ne k$. Thus, the upper bound of the overall secrecy capacity considering all the actuators is derived as ${\overset{\lower0.5em\hbox{$\smash{\scriptscriptstyle\smile}$}}{C} ^{\sec }}{\text{ = }}\overset{\lower0.5em\hbox{$\smash{\scriptscriptstyle\smile}$}}{C} _k^{{\text{sec}}} + \sum\limits_{i = 1,i \ne k}^K {{C_l}} $.
The total power consumed by the gateway controller is modeled as \cite{HZhang14_24,Zappone16_26,Zi16}

\[{P_{{\text{total}}}}{\text{ = }}\frac{1}{\alpha }\left( {\sum\limits_{k = 1}^K {{P_k}}  + {P_{\text{n}}}} \right) + {N_{{\text{RF}}}}\left( {{P_{{\text{RF}}}} + {N_{{\text{TX}}}}{P_{{\text{PS}}}}} \right) + {P_{\text{C}}},\tag{6}\]
where $\alpha $ is the efficiency of the power amplifier, and the term $\left( {\sum\limits_{k = 1}^K {{P_k}}  + {P_{\text{n}}}} \right)$ denotes the power consumed to transmit the signals and AN. ${P_{{\text{RF}}}}$ and ${P_{{\text{PS}}}}$ denotes the power consumed by each RF chain and each phase shifter, respectively.   is the constant power consumption at the gateway controller.

Based on the established secrecy capacity and power consumption model, the lower bound of the secure energy efficiency is expressed as

\[\overset{\lower0.5em\hbox{$\smash{\scriptscriptstyle\smile}$}}{\eta } {\text{ = }}\frac{{W\left( {\overset{\lower0.5em\hbox{$\smash{\scriptscriptstyle\smile}$}}{C} _k^{{\text{sec}}} + \sum\limits_{i = 1,i \ne k}^K {{C_l}} } \right)}}{{\frac{1}{\alpha }\left( {\sum\limits_{k = 1}^K {{P_k}}  + {P_{\text{n}}}} \right) + {N_{{\text{RF}}}}\left( {{P_{{\text{RF}}}} + {N_{{\text{TX}}}}{P_{{\text{PS}}}}} \right) + {P_{\text{C}}}}}.\tag{7}\]

In this paper, the goal is to maximize the secure energy efficiency of the gateway controller in the downlink transmission when the transmit power is constrained, the minimum secrecy capacity is required, and the hardware limitations are considered. The maximization procedure is implemented by optimization the gateway controller baseband precoding matrices ${{\mathbf{B}}_{\text{s}}}$ and ${{\mathbf{B}}_{\text{n}}}$, as well as the RF precoding matrices ${{\mathbf{F}}_{\text{s}}}$ and ${{\mathbf{F}}_{\text{n}}}$. The optimization problem is formulated as

\[\begin{gathered}
  \left( {{\mathbf{F}}_{\text{s}}^{{\text{opt}}},{\mathbf{B}}_{\text{s}}^{{\text{opt}}},{\mathbf{F}}_{\text{n}}^{{\text{opt}}},{\mathbf{B}}_{\text{n}}^{{\text{opt}}}} \right) = \mathop {{\text{arg}}\;\max }\limits_{{{\mathbf{F}}_{\text{s}}},{{\mathbf{B}}_{\text{s}}},{{\mathbf{F}}_{\text{n}}},{{\mathbf{B}}_{\text{n}}}} \;\overset{\lower0.5em\hbox{$\smash{\scriptscriptstyle\smile}$}}{\eta }  \hfill \\
  \;\;\;\;\;\;\;\;\;\;\;\;\;\;\;\;\;\;s.\;t.\;\;\sum\limits_{k = 1}^K {{P_k}}  + {P_{\text{n}}} \leqslant {P_{\max }} \hfill \\
  \;\;\;\;\;\;\;\;\;\;\;\;\;\;\;\;\;\;\;\;\;\;\;\;{{\overset{\lower0.5em\hbox{$\smash{\scriptscriptstyle\smile}$}}{C} }^{\sec }} > C_0^{\sec } \hfill \\
  \;\;\;\;\;\;\;\;\;\;\;\;\;\;\;\;\;\;\;\;\;\;\;\;{\left| {{{\left[ {{{\mathbf{F}}_{\text{s}}}} \right]}_{i,j}}} \right|^2} = {\left| {{{\left[ {{{\mathbf{F}}_{\text{n}}}} \right]}_{i,j}}} \right|^2} = \frac{1}{{{N_{{\text{TX}}}}}} \hfill \\
\end{gathered} ,\tag{8}\]
where ${P_{\max }}$ is the maximum transmit power at the gateway controller, $C_0^{\sec }$ the minimum secrecy capacity required to satisfy the quality of service (QoS) constraint. The constraint $\;{\left| {{{\left[ {{{\mathbf{F}}_{\text{s}}}} \right]}_{i,j}}} \right|^2} = {\left| {{{\left[ {{{\mathbf{F}}_{\text{n}}}} \right]}_{i,j}}} \right|^2} = \frac{1}{{{N_{{\text{TX}}}}}}$ is due to that analog precoding is implemented by phase shifters, which can only changes the phases of the signals. Thus, the amplitudes of the entries of the analog precoding matrices are constant and configured as $\frac{1}{{{N_{{\text{TX}}}}}}$ without loss of generality.

\section{Problem formulation}
\label{sec4}
\subsection{RF precoding design}

It is intractable to obtain the global optimal solution for the optimization problem in (8) because of the non-convexity of the objective function and the feasible domain. The coupling of the precoding matrices makes it even more difficult to find the optimized result. To tackle this problem, we seek to directly design the RF and baseband precoders, then try to solve the secure energy efficiency problem with a tractable form.

Firstly assume the analog RF preocders of AN and downlink signals are identical, i.e. ${\mathbf{F}}{\text{ = }}{{\mathbf{F}}_{\text{n}}}{\text{ = }}{{\mathbf{F}}_{\text{s}}}$ \cite{Zhu17}. When the downlink signals and AN have identical RF precoders ${\mathbf{F}}$, the RF precoder utilizes the all the degrees of freedom provided by the RF chains to enhance the actuator received signals and eavesdropper received AN. In this case, the highest secrecy capacity is achieved \cite[Proposition 2]{Zhu17}. Similar to \cite{Liang14}, further configure the RF precoder as

\[{\mathbf{F}}{\text{ = }}\left\{ \begin{gathered}
  \frac{1}{{\sqrt {{N_{{\text{TX}}}}} }}{e^{j{\phi _{i,j}}}},\;{\text{for}}\;1 \leqslant j \leqslant K \hfill \\
  \frac{1}{{\sqrt {{N_{{\text{TX}}}}} }}{e^{j{\varphi _{i,j}}}},\;{\text{for}}\;K + 1 \leqslant j \leqslant {N_{{\text{RF}}}} \hfill \\
\end{gathered}  \right.,\tag{9}\]
where ${\phi _{i,j}}$ is the phase of the $\left( {i,j} \right){\text{th}}$ entry of the conjugate transpose of the downlink channel matrix for $K$ ANs, i.e. the $\left( {i,j} \right){\text{th}}$ entry of the matrix ${\mathbf{H}} = \left[ {{{\mathbf{h}}_1},...,{{\mathbf{h}}_k}{\text{,}}...{\text{,}}{{\mathbf{h}}_K}} \right]$. ${\varphi _{i,j}}$ is configured as a uniformly distributed variable within $\left. {\left[ {0,2\pi } \right.} \right)$, i.e. ${\varphi _{i,j}} \sim \mathcal{U}\left. {\left[ {0,2\pi } \right.} \right)$. The configuration is based on the assumption that the gateway controller has full knowledge of the channels.

Combine the RF precoding matrix with the downlink channel matrices, then the equivalent downlink channel vectors and matrices of the $k{\text{th}}$ actuator, $K$ actuators and the eavesdropper are given by ${\mathbf{h}}_{{\text{eq}},k}^H = {\mathbf{h}}_k^H{\mathbf{F}}$, ${\mathbf{H}}_{{\text{eq}}}^H{\text{ = }}{\mathbf{H}}_{{\text{eq}}}^H{\mathbf{F}}$ and ${\mathbf{H}}_{{\text{eq,E}}}^H = {\mathbf{H}}_{\text{E}}^H{\mathbf{F}}$, respectively.

The received SINR of the $k{\text{th}}$ actuator in (3) is transformed as

\[{\gamma _k} = \frac{{{\mathbf{h}}_{{\text{eq}},k}^H{{\mathbf{b}}_{{\text{s,}}k}}{\mathbf{b}}_{{\text{s}},k}^H{{\mathbf{h}}_{{\text{eq}},k}}}}{{\sum\limits_{i = 1,i \ne k}^K {{\mathbf{h}}_{{\text{eq}},k}^H{{\mathbf{b}}_{{\text{s,}}i}}{\mathbf{b}}_{{\text{s}},i}^H{{\mathbf{h}}_{{\text{eq}},k}}}  + {\mathbf{h}}_{{\text{eq}},k}^H{{\mathbf{B}}_{\text{n}}}{\mathbf{B}}_{\text{n}}^H{{\mathbf{h}}_{{\text{eq}},k}} + \sigma _w^2}}.\tag{10}\]

The upper-bound of the received SINR at the eavesdropper in (4) is transformed as：

\[{\overset{\lower0.5em\hbox{$\smash{\scriptscriptstyle\frown}$}}{\gamma } _{{\text{E}},k}} = {\mathbf{b}}_{{\text{s,}}k}^H{{\mathbf{H}}_{{\text{eq,E}}}}{\left( {{\mathbf{H}}_{{\text{eq,E}}}^H{{\mathbf{B}}_{\text{n}}}{\mathbf{B}}_{\text{n}}^H{{\mathbf{H}}_{{\text{eq,E}}}}} \right)^{ - 1}}{\mathbf{H}}_{{\text{eq,E}}}^H{{\mathbf{b}}_{{\text{s,}}k}}.\tag{11}\]

We further assume ${P'_{\text{C}}} = {N_{{\text{RF}}}}\left( {{P_{{\text{RF}}}} + {N_{{\text{TX}}}}{P_{{\text{PS}}}}} \right) + {P_{\text{C}}}$, then the lower-bound of the secure energy efficiency is rewritten as

\[\overset{\lower0.5em\hbox{$\smash{\scriptscriptstyle\smile}$}}{\eta } {\text{ = }}\frac{{W\left[ {\sum\limits_{i = 1}^K {{{\log }_2}\left( {1 + {\gamma _k}} \right)} } \right.\left. { - {{\log }_2}\left( {1 + {{\overset{\lower0.5em\hbox{$\smash{\scriptscriptstyle\frown}$}}{\gamma } }_{{\text{E}},k}}} \right)} \right]}}{{\frac{1}{\alpha }\left( {\sum\limits_{k = 1}^K {{P_k}}  + {P_{\text{n}}}} \right) + {{P'}_{\text{C}}}}}.\tag{12}\]

\subsection{Baseband precoding design}
Assume the baseband precoder for the signals to $K$ actuator as zero precoding (ZF) precoders \cite{Zhu17,Liang14}, i.e.

\[{{\mathbf{B}}_{\text{s}}} = {{\mathbf{H}}_{{\text{eq}}}}{\left( {{\mathbf{H}}_{{\text{eq}}}^H{{\mathbf{H}}_{{\text{eq}}}}} \right)^{ - 1}}{\mathbf{D}},\tag{13}\]

Where ${\mathbf{D}}$ is the $K \times K$ diagonal normalization matrix, whose $k{\text{th}}$ diagonal entry is denoted as ${\zeta _k}$. To ensure ${P_k} = {\left\| {{\mathbf{F}}{{\mathbf{b}}_{{\text{s,}}k}}} \right\|^2}$ with the configured RF precoding matrix in (10), it is easy to derive ${\zeta _k}$ as [34][35]

\[{\zeta _k}{\text{ = }}\sqrt {\frac{{\pi {P_k}}}{4}\left( {{N_{{\text{Tx}}}} - 1} \right)} .\tag{14}\]

To suppress to received signal by eavesdropper while keeping the signal received by actuator little degraded, the artificial noise is aligned in the null space of the $K$ actuators' downlink channel \cite{Wang15_20,HMWang15_21}. The singular value decomposition (SVD) of the actuators' equivalent downlink channel is written as

\[{\mathbf{H}}_{{\text{eq}}}^H{\text{ = }}{\mathbf{U\Sigma }}{{\mathbf{V}}^H}.\tag{15}\]
where ${\mathbf{U}} \in {\mathbb{C}^{K \times K}}$ consists of $K$ left singular vectors, ${\mathbf{\Sigma }} \in {\mathbb{C}^{K \times {N_{{\text{RF}}}}}}$ is the diagonal matrix which consists of K singular values, and ${{\mathbf{V}}^H} \in {\mathbb{C}^{{N_{{\text{RF}}}} \times {N_{{\text{RF}}}}}}$ consists of ${N_{{\text{RF}}}}$ right singular vectors. The last ${N_{{\text{RF}}}} - K$ columns of ${\mathbf{V}}$ corresponds to zero singular values of ${\mathbf{H}}_{{\text{eq}}}^H$. To ensure ${\mathbf{H}}_{{\text{eq}}}^H{{\mathbf{B}}_{\text{n}}}{\text{ = }}{\mathbf{0}}$, the baseband precoding matrix for the AN to the eavesdropper is then designed as

\[{{\mathbf{B}}_{\text{n}}}{\text{ = }}{\mathbf{V}}\left( {:,K + 1:{N_{{\text{RF}}}}} \right).\tag{16}\]

\subsection{Secure energy efficiency optimization}
Substituting (13)-(16) into (10), the received SINR of the $k{\text{th}}$ actuator is derived as

\[{\gamma _k} = \frac{{\zeta _k^2}}{{\sigma _w^2}} = \frac{{\pi {P_k}}}{{4\sigma _w^2}}\left( {{N_{{\text{Tx}}}} - 1} \right).\tag{17}\]

The capacity of the $k{\text{th}}$ actuator is then obtained as

\[{C_k} = {\log _2}\left[ {1 + \frac{{\pi {P_k}}}{{4\sigma _w^2}}\left( {{N_{{\text{Tx}}}} - 1} \right)} \right].\tag{18}\]

Based on the massive MIMO configurations, the upper-bound of the capacity of the eavesdropper is derived as \cite[Theorem 2]{Zhu14}

\[{\overset{\lower0.5em\hbox{$\smash{\scriptscriptstyle\frown}$}}{C} _{{\text{E}},k}} = {\log _2}\left( {1 + {{\overset{\lower0.5em\hbox{$\smash{\scriptscriptstyle\frown}$}}{\gamma } }_{{\text{E}},k}}} \right) \approx {\log _2}\left( {1 + \frac{{M\sum\limits_{k = 1}^K {{P_k}} }}{{{P_{\text{n}}}\left( {1 - \frac{M}{{{N_{{\text{RF}}}} - K}}} \right)}}} \right).\tag{19}\]

Substitute (18) and (19) into (8), the lower-bound of the secure energy efficiency can be written as

\[\overset{\lower0.5em\hbox{$\smash{\scriptscriptstyle\smile}$}}{\eta }  \approx \frac{{W{{\overset{\lower0.5em\hbox{$\smash{\scriptscriptstyle\smile}$}}{C} }^{\sec }}\left( {{\mathbf{P}},\;{P_{\text{n}}}} \right)}}{{{P_{{\text{total}}}}\left( {{\mathbf{P}},\;{P_{\text{n}}}} \right)}},\tag{20a}\]
where, ${\mathbf{P}} = \left[ {{P_1},...,{P_k},...{P_K}} \right]$, and

\[{\overset{\lower0.5em\hbox{$\smash{\scriptscriptstyle\smile}$}}{C} ^{\sec }}\left( {{\mathbf{P}},\;{P_{\text{n}}}} \right) = {C_{\text{s}}}\left( {\mathbf{P}} \right) - {\overset{\lower0.5em\hbox{$\smash{\scriptscriptstyle\frown}$}}{C} _{{\text{E}},k}}\left( {{\mathbf{P}},\;{P_{\text{n}}}} \right),\tag{20b}\]

\[{C_{\text{s}}}\left( {\mathbf{P}} \right) = \sum\limits_{i = 1}^K {{{\log }_2}\left( {1 + \frac{{\pi {P_k}\left( {{N_{{\text{Tx}}}} - 1} \right)}}{{4\sigma _w^2}}} \right)} ,\tag{20c}\]

\[{\overset{\lower0.5em\hbox{$\smash{\scriptscriptstyle\frown}$}}{C} _{{\text{E}},k}}\left( {{\mathbf{P}},\;{P_{\text{n}}}} \right) = {\log _2}\left( {1 + \frac{{M\sum\limits_{k = 1}^K {{P_k}} }}{{K{P_{\text{n}}}\left( {1 - \frac{M}{{{N_{{\text{RF}}}} - K}}} \right)}}} \right),\tag{20d}\]

\[{P_{{\text{total}}}}\left( {{\mathbf{P}},\;{P_{\text{n}}}} \right) = \frac{1}{\alpha }\left( {\sum\limits_{k = 1}^K {{P_k}}  + {P_{\text{n}}}} \right) + {P'_{\text{C}}},\tag{20e}\]

Then the secure energy efficiency optimization problem is rewritten as：

\[\begin{gathered}
  \left( {P_{\text{s}}^{{\text{opt}}},P_{\text{n}}^{{\text{opt}}}} \right) = \mathop {{\text{arg}}\;\max }\limits_{{\mathbf{P}},\;{P_{\text{n}}}} \;\overset{\lower0.5em\hbox{$\smash{\scriptscriptstyle\smile}$}}{\eta }  \approx \frac{{W{{\overset{\lower0.5em\hbox{$\smash{\scriptscriptstyle\smile}$}}{C} }^{\sec }}\left( {{\mathbf{P}},\;{P_{\text{n}}}} \right)}}{{{P_{{\text{total}}}}\left( {{\mathbf{P}},\;{P_{\text{n}}}} \right)}} \hfill \\
  \;\;\;\;\;\;\;\;\;\;\;\;\;s.\;t.\;\;\sum\limits_{k = 1}^K {{P_k}}  + {P_{\text{n}}} \leqslant {P_{\max }} \hfill \\
  \;\;\;\;\;\;\;\;\;\;\;\;\;\;\;\;\;\;\;\;{{\overset{\lower0.5em\hbox{$\smash{\scriptscriptstyle\smile}$}}{C} }^{\sec }} > C_0^{\sec } \hfill \\
\end{gathered} .\tag{21}\]

Although only two constraints exist in (21), the objective function and feasible domain in (21) are still non-convex. It is still unable to directly solve the optimization problem with a standard method. Noting that the objective function in (21) has a fractional form, the optimization problem can be transformed into a parameterized polynomial subtractive form:

\[\left( {{{\mathbf{P}}^{{\text{opt}}}},P_{\text{n}}^{{\text{opt}}}} \right) = \mathop {\arg \;max}\limits_{\left( {{\mathbf{P}},\;{P_{\text{n}}}} \right) \in \mathbb{Q}} \;\left[ \begin{gathered}
  W{{\overset{\lower0.5em\hbox{$\smash{\scriptscriptstyle\smile}$}}{C} }^{\sec }}\left( {{\mathbf{P}},\;{P_{\text{n}}}} \right) \hfill \\
   - \overset{\lower0.5em\hbox{$\smash{\scriptscriptstyle\smile}$}}{\eta } {P_{{\text{total}}}}\left( {{\mathbf{P}},\;{P_{\text{n}}}} \right) \hfill \\
\end{gathered}  \right],\tag{22a}\]
where

\[\mathbb{Q} = \left[ {\left( {{\mathbf{P}},\;{P_{\text{n}}}} \right):{{\overset{\lower0.5em\hbox{$\smash{\scriptscriptstyle\smile}$}}{C} }^{\sec }}\left( {{\mathbf{P}},\;{P_{\text{n}}}} \right) > C_0^{\sec },\sum\limits_{k = 1}^K {{P_k}}  + {P_{\text{n}}} \leqslant {P_{\max }}} \right].\tag{22b}\]

(22) is a parametric programming problem with parameter $\overset{\lower0.5em\hbox{$\smash{\scriptscriptstyle\smile}$}}{\eta } $. According to Dinkelbach's method \cite{Dinkelbach67}, (22) can be solved by iteratively solving the following problem:

\[\left( {{\mathbf{P}}_{\left( i \right)}^{{\text{opt}}},P_{{\text{n,}}\left( i \right)}^{{\text{opt}}}} \right) = \mathop {\arg \;max}\limits_{\left( {{\mathbf{P}},\;{P_{\text{n}}}} \right) \in \mathbb{Q}} \;\left[ \begin{gathered}
  W{{\overset{\lower0.5em\hbox{$\smash{\scriptscriptstyle\smile}$}}{C} }^{{\text{sec}}}}\left( {{\mathbf{P}},\;{P_{\text{n}}}} \right) \hfill \\
   - {{\overset{\lower0.5em\hbox{$\smash{\scriptscriptstyle\smile}$}}{\eta } }_{\left( i \right)}}{P_{{\text{total}}}}\left( {{\mathbf{P}},\;{P_{\text{n}}}} \right) \hfill \\
\end{gathered}  \right],\tag{23}\]
where ${\overset{\lower0.5em\hbox{$\smash{\scriptscriptstyle\smile}$}}{\eta } _{\left( i \right)}}$ is the parameter at the $i{\text{th}}$ iteration. An initial value ${\overset{\lower0.5em\hbox{$\smash{\scriptscriptstyle\smile}$}}{\eta } _{\left( 0 \right)}}$ is preset at the beginning of the iteration. The optimized transmit power ${\mathbf{P}}_{\left( i \right)}^{{\text{opt}}}$ and $P_{{\text{n,}}\left( i \right)}^{{\text{opt}}}$ is obtained from solving (23). Based on Dinkelbach's method, the termination condition for the iteration process is expressed as

\[\left| {W{{\overset{\lower0.5em\hbox{$\smash{\scriptscriptstyle\smile}$}}{C} }^{{\text{sec}}}}\left( {{\mathbf{P}}_{\left( i \right)}^{{\text{opt}}},P_{{\text{n,}}\left( i \right)}^{{\text{opt}}}} \right) - {{\overset{\lower0.5em\hbox{$\smash{\scriptscriptstyle\smile}$}}{\eta } }_{\left( i \right)}}{P_{{\text{total}}}}\left( {{\mathbf{P}}_{\left( i \right)}^{{\text{opt}}},P_{{\text{n,}}\left( i \right)}^{{\text{opt}}}} \right)} \right| \leqslant \upsilon ,\tag{24}\]
where $\upsilon $ is the convergence tolerance factor with $\upsilon  \geqslant 0$. If (24) is satisfied, ${\mathbf{P}}_{\left( i \right)}^{{\text{opt}}}$ and $P_{{\text{n,}}\left( i \right)}^{{\text{opt}}}$ are the optimal transmit power for the secure EE optimization problem in (22) and (21). Otherwise, ${\overset{\lower0.5em\hbox{$\smash{\scriptscriptstyle\smile}$}}{\eta } _{\left( {i + 1} \right)}}$ is calculated as follows for the next iteration

\[{\overset{\lower0.5em\hbox{$\smash{\scriptscriptstyle\smile}$}}{\eta } _{\left( {i + 1} \right)}} = \frac{{W{{\overset{\lower0.5em\hbox{$\smash{\scriptscriptstyle\smile}$}}{C} }^{{\text{sec}}}}\left( {{\mathbf{P}}_{\left( i \right)}^{{\text{opt}}},P_{{\text{n,}}\left( i \right)}^{{\text{opt}}}} \right)}}{{{P_{{\text{total}}}}\left( {{\mathbf{P}}_{\left( i \right)}^{{\text{opt}}},P_{{\text{n,}}\left( i \right)}^{{\text{opt}}}} \right)}}.\tag{25}\]

The convergence of the above iteration process is proved in \cite{Dinkelbach67,Schaible76,Ng12}. For the limitation on the length of the paper, the detailed proofs are neglected here.

However, directly solving (23) is still intractable because of the non-convexity of the feasible domain $\mathbb{Q}$. Then the penalty function method \cite{Bazaraa13,Dinh14} is utilized to transform the feasible domain and objective function in (23) into a more tractable form:

\[\begin{gathered}
  \left[ {{\mathbf{P}}\left( {{\omega _j}} \right),{P_{\text{n}}}\left( {{\omega _j}} \right)} \right] \hfill \\
  \;\;\;\;\;\;\;\; = \mathop {\arg \;\min }\limits_{\left( {{\mathbf{P}},\;{P_{\text{n}}}} \right) \in \mathbb{Q}'} \;\left\{ \begin{gathered}
  {{\overset{\lower0.5em\hbox{$\smash{\scriptscriptstyle\smile}$}}{\eta } }_{\left( i \right)}}{P_{{\text{total}}}}\left( {{\mathbf{P}},\;{P_{\text{n}}}} \right) - W{{\overset{\lower0.5em\hbox{$\smash{\scriptscriptstyle\smile}$}}{C} }^{{\text{sec}}}}\left( {{\mathbf{P}},\;{P_{\text{n}}}} \right) \hfill \\
   + {\omega _j}\left[ {t + W{{\overset{\lower0.5em\hbox{$\smash{\scriptscriptstyle\frown}$}}{C} }_{{\text{E}},k}}\left( {{\mathbf{P}},\;{P_{\text{n}}}} \right)} \right] \hfill \\
\end{gathered}  \right\} \hfill \\
\end{gathered},\tag{26a}\]

where

\[\mathbb{Q}' = \left[ \begin{gathered}
  \left( {{\mathbf{P}},\;{P_{\text{n}}}} \right):C_0^{\sec } - {C_{\text{s}}}\left( {{\mathbf{P}},{P_{\text{n}}}} \right) \leqslant t, \hfill \\
   - {{\overset{\lower0.5em\hbox{$\smash{\scriptscriptstyle\frown}$}}{C} }_{{\text{E}},k}}\left( {{\mathbf{P}},\;{P_{\text{n}}}} \right) \leqslant t, \hfill \\
  \sum\limits_{k = 1}^K {{P_k}}  + {P_{\text{n}}} \leqslant {P_{\max }} \hfill \\
\end{gathered}  \right].\tag{26b}\]

The problem in (26) is also iteratively solved. ${\omega _j}$ is the penalty factor at the $j{\text{th}}$ step. The penalty factor begins with an initial value ${\omega _0}$. $t$ is an auxiliary variable. Based on the penalty function method, iteratively solving (27) with an increasing penalty factor is equivalent to solving the optimization problem in (26). At the $j{\text{th}}$ step of solving (26), the power ${\mathbf{P}}\left( {{\omega _j}} \right)$ and ${P_{\text{n}}}\left( {{\omega _j}} \right)$ are obtained. If either of the termination conditions

\[{\omega _j}\max \left\{ {C_0^{\sec } - {{\overset{\lower0.5em\hbox{$\smash{\scriptscriptstyle\smile}$}}{C} }^{{\text{sec}}}}\left[ {{\mathbf{P}}\left( {{\omega _j}} \right),{P_{\text{n}}}\left( {{\omega _j}} \right)} \right],0} \right\} \leqslant \rho ,\tag{27a}\]

or

\[j \geqslant J\tag{27b}\]
is satisfied, the iteration process stops. And the obtained power ${\mathbf{P}}\left( {{\omega _j}} \right)$ and ${P_{\text{n}}}\left( {{\omega _j}} \right)$ are the optimized power in (23), i.e. ${\mathbf{P}}_{\left( i \right)}^{{\text{opt}}} = {\mathbf{P}}\left( {{\omega _j}} \right)$，$P_{{\text{n,}}\left( i \right)}^{{\text{opt}}} = {P_{\text{n}}}\left( {{\omega _j}} \right)$. $\rho $ and $J$ are the convergence threshold and maximum iteration number in (27), respectively. If neither condition is met, the iteration continues with updating the penalty factor ${\omega _{j + 1}} = \mu {\omega _j}$, where $\mu $ is the iteration increment factor with $\mu  > 1$. The convergence of the above iteration process is guaranteed. For detailed verifications, the readers can refer to \cite{DWang16_23,Dinh14}.

To further solve the minimization problem in (26), (26a) is rewritten in the flowing form


\[\begin{gathered}
  \left[ {{\mathbf{P}}\left( {{\omega _j}} \right),{P_{\text{n}}}\left( {{\omega _j}} \right)} \right] \hfill \\
  \;\;\;\;\;\;\;\; = \mathop {\arg \;\min }\limits_{\left( {{\mathbf{P}},\;{P_{\text{n}}}} \right) \in \mathbb{Q}'} \;\left\{ {{\Gamma _1}\left( {{\mathbf{P}},\;{P_{\text{n}}},{\omega _j}} \right) - {\Gamma _2}\left( {{\mathbf{P}},\;{P_{\text{n}}},{\omega _j}} \right)} \right\} \hfill \\
\end{gathered},\tag{28a} \]
where

\[{\Gamma _1}\left( {{\mathbf{P}},\;{P_{\text{n}}},{\omega _j}} \right) = {\overset{\lower0.5em\hbox{$\smash{\scriptscriptstyle\smile}$}}{\eta } _{\left( i \right)}}{P_{{\text{total}}}}\left( {{\mathbf{P}},\;{P_{\text{n}}}} \right) - W{C_{\text{s}}}\left( {\mathbf{P}} \right) + {\omega _j}t,\tag{28b}\]

\[{\Gamma _2}\left( {{\mathbf{P}},\;{P_{\text{n}}},{\omega _j}} \right) =  - W\left( {{\omega _j} + 1} \right){\overset{\lower0.5em\hbox{$\smash{\scriptscriptstyle\frown}$}}{C} _{{\text{E}},k}}\left( {{\mathbf{P}},\;{P_{\text{n}}}} \right),\tag{28c}\]

Based on the expressions of ${P_{{\text{total}}}}\left( {{\mathbf{P}},\;{P_{\text{n}}}} \right)$, ${C_{\text{s}}}\left( {\mathbf{P}} \right)$ and ${C_{\text{n}}}\left( {{\mathbf{P}},\;{P_{\text{n}}}} \right)$ in (6) and (21), it is easy to find that ${\Gamma _1}\left( {{\mathbf{P}},\;{P_{\text{n}}},{\omega _j}} \right)$ is an affine function with respect to ${P_{\text{n}}}$ when ${\mathbf{P}}$ is fixed. And because of the assumption that the eavesdropper is equipped with an enough number of antennas to ensure $M > {N_{{\text{RF}}}} - K$, ${\Gamma _2}\left( {{\mathbf{P}},\;{P_{\text{n}}},{\omega _j}} \right)$ is a convex function with respect to ${P_{\text{n}}}$ when ${\mathbf{P}}$ is fixed. When ${P_{\text{n}}}$ is fixed, ${\Gamma _1}\left( {{\mathbf{P}},\;{P_{\text{n}}},{\omega _j}} \right)$ and ${\Gamma _2}\left( {{\mathbf{P}},\;{P_{\text{n}}},{\omega _j}} \right)$ are convex functions with respect to ${\mathbf{P}}$. As a result, it can be concluded that the when either of ${P_{\text{n}}}$ or ${\mathbf{P}}$ is fixed, the optimization problem in (28) is a standard difference of convex functions programming problem, i.e. the DC programming problem \cite{Phan12}. The alternate search method can be employed to iteratively solve (28). At each iteration during the alternate search, only one of ${P_{\text{n}}}$ or ${\mathbf{P}}$ is optimized by the DC programming method while the other is fixed \cite{Phan12}. To be specific, at the $m{\text{th}}$ step of the alternate search, ${{\mathbf{P}}^{\left( {m + 1} \right)}}\left( {{\omega _j}} \right)$ is obtained by the DC programming method with the fixed $P_{\text{n}}^{\left( m \right)}\left( {{\omega _j}} \right)$

\[{{\mathbf{P}}^{\left( {m + 1} \right)}}\left( {{\omega _j}} \right) = \mathop {\arg \;\min }\limits_{{\mathbf{P}} \in {{\mathbb{Q}'}_{\mathbf{P}}}} \;\left\{ \begin{gathered}
  {\Gamma _1}\left[ {{\mathbf{P}},P_{\text{n}}^{\left( m \right)}\left( {{\omega _j}} \right),{\omega _j}} \right] \hfill \\
   - {\Gamma _2}\left[ {{\mathbf{P}},P_{\text{n}}^{\left( m \right)}\left( {{\omega _j}} \right),{\omega _j}} \right] \hfill \\
\end{gathered}  \right\},\tag{29a}\]
where

\[{\mathbb{Q}'_{\mathbf{P}}} = \left[ \begin{gathered}
  {\mathbf{P}}:C_0^{\sec } - {C_{\text{s}}}\left( {{\mathbf{P}},\;P_{\text{n}}^{\left( m \right)}\left( {{\omega _j}} \right)} \right) \leqslant t, \hfill \\
   - {{\overset{\lower0.5em\hbox{$\smash{\scriptscriptstyle\frown}$}}{C} }_{{\text{E}},k}}\left( {{\mathbf{P}},\;P_{\text{n}}^{\left( m \right)}\left( {{\omega _j}} \right)} \right) \leqslant t, \hfill \\
  \sum\limits_{k = 1}^K {{P_k}}  + P_{\text{n}}^{\left( m \right)}\left( {{\omega _j}} \right) \leqslant {P_{\max }} \hfill \\
\end{gathered}  \right].\tag{29b}\]

Then $P_{\text{n}}^{\left( {m + 1} \right)}\left( {{\omega _j}} \right)$ is obtained by the DC programming method with the fixed ${{\mathbf{P}}^{\left( {m + 1} \right)}}\left( {{\omega _j}} \right)$, i.e.

\[P_{\text{n}}^{\left( {m + 1} \right)}\left( {{\omega _j}} \right) = \mathop {\arg \;\min }\limits_{{P_{\text{n}}} \in {{\mathbb{Q}'}_{{P_{\text{n}}}}}} \;\left\{ \begin{gathered}
  {\Gamma _1}\left[ {{{\mathbf{P}}^{\left( {m + 1} \right)}}\left( {{\omega _j}} \right),\;{P_{\text{n}}},{\omega _j}} \right] \hfill \\
   - {\Gamma _2}\left[ {{{\mathbf{P}}^{\left( {m + 1} \right)}}\left( {{\omega _j}} \right),\;{P_{\text{n}}},{\omega _j}} \right] \hfill \\
\end{gathered}  \right\},\tag{30a}\]
where

\[{\mathbb{Q}'_{{P_n}}} = \left[ \begin{gathered}
  {P_n}:C_0^{\sec } - {C_{\text{s}}}\left( {{{\mathbf{P}}^{\left( {m + 1} \right)}}\left( {{\omega _j}} \right),\;{P_n}} \right) \leqslant t, \hfill \\
   - {{\overset{\lower0.5em\hbox{$\smash{\scriptscriptstyle\frown}$}}{C} }_{{\text{E}},k}}\left( {{{\mathbf{P}}^{\left( {m + 1} \right)}}\left( {{\omega _j}} \right),\;{P_n}} \right) \leqslant t, \hfill \\
  \sum\limits_{k = 1}^K {P_k^{\left( {m + 1} \right)}} \left( {{\omega _j}} \right) + {P_n} \leqslant {P_{\max }} \hfill \\
\end{gathered}  \right].\tag{30b}\]

Solving (29) and (30) with DC programing is quite straightforward \cite{Phan12,Chen13}. The detailed process is neglected in this paper.

Denoting

\[\begin{gathered}
  {\Upsilon _m} = {\Gamma _1}\left( {{\mathbf{P}}_{\left( i \right)}^{\left( {m + 1} \right)},\;P_{{\text{n,}}\left( i \right)}^{\left( {m + 1} \right)}} \right) - {\Gamma _2}\left( {{\mathbf{P}}_{\left( i \right)}^{\left( {m + 1} \right)},\;P_{{\text{n,}}\left( i \right)}^{\left( {m + 1} \right)}} \right) \hfill \\
  \;\;\;\;\;\;\;\;\; - \left[ {{\Gamma _1}\left( {{\mathbf{P}}_{\left( i \right)}^{\left( m \right)},\;P_{{\text{n,}}\left( i \right)}^{\left( m \right)}} \right) - {\Gamma _2}\left( {{\mathbf{P}}_{\left( i \right)}^{\left( m \right)},\;P_{{\text{n,}}\left( i \right)}^{\left( m \right)}} \right)} \right] \hfill \\
\end{gathered} ,\tag{31}\]
if either of the termination conditions $\left| {{\Upsilon _m}} \right| \leqslant \rho '$ or $m + 1 > J'$ is satisfied at the end of the $m{\text{th}}$ step of the alternate search, the alternate search terminates. $\rho '$ and $J'$ are the convergence threshold and maximum iteration number for the iteration process in (29) and (30), respectively. The optimization problems in (28) and (26) are solved with ${\mathbf{P}}\left( {{\omega _j}} \right) = {{\mathbf{P}}^{\left( {m + 1} \right)}}\left( {{\omega _j}} \right)$ and ${P_{\text{n}}}\left( {{\omega _j}} \right) = P_{\text{n}}^{\left( {m + 1} \right)}\left( {{\omega _j}} \right)$. The detailed verifications for the convergence of the alternate search is given in \cite{DWang16_23}, which is not listed here.

\begin{algorithm*}[!t]
\caption{\textbf{Secure Energy Efficient Hybrid Precoding (SEEHP) algorithm}.}
\label{alg1}
\begin{algorithmic}
\STATE \textbf{Begin:} \begin{enumerate}

\item Preset $i = 0$, $t$, $\upsilon $, ${\omega _0} > \omega $, ${\overset{\lower0.5em\hbox{$\smash{\scriptscriptstyle\smile}$}}{\eta } _{\left( 0 \right)}} = \frac{{W{{\overset{\lower0.5em\hbox{$\smash{\scriptscriptstyle\smile}$}}{C} }^{{\text{sec}}}}\left( {{\mathbf{P}}_{\left( 0 \right)}^{{\text{opt}}},P_{{\text{n,}}\left( 0 \right)}^{{\text{opt}}}} \right)}}{{{P_{{\text{total}}}}\left( {{\mathbf{P}}_{\left( 0 \right)}^{{\text{opt}}},P_{{\text{n,}}\left( 0 \right)}^{{\text{opt}}}} \right)}}$;

\item While $\left| {W{{\overset{\lower0.5em\hbox{$\smash{\scriptscriptstyle\smile}$}}{C} }^{{\text{sec}}}}\left( {{\mathbf{P}}_{\left( i \right)}^{{\text{opt}}},P_{{\text{n,}}\left( i \right)}^{{\text{opt}}}} \right) - {{\overset{\lower0.5em\hbox{$\smash{\scriptscriptstyle\smile}$}}{\eta } }_{\left( i \right)}}{P_{{\text{total}}}}\left( {{\mathbf{P}}_{\left( i \right)}^{{\text{opt}}},P_{{\text{n,}}\left( i \right)}^{{\text{opt}}}} \right)} \right| \leqslant \upsilon $

\item \qquad preset  $j = 0$, ${\mathbf{P}}\left( {{\omega _0}} \right) = {\mathbf{P}}_{\left( i \right)}^{{\text{opt}}}$,
${P_{\text{n}}}\left( {{\omega _0}} \right) = P_{{\text{n,}}\left( i \right)}^{{\text{opt}}}$;

\item \qquad While ${\omega _j}\max \left\{ {C_0^{\sec } - {{\overset{\lower0.5em\hbox{$\smash{\scriptscriptstyle\smile}$}}{C} }^{{\text{sec}}}}\left[ {{\mathbf{P}}\left( {{\omega _j}} \right),{P_{\text{n}}}\left( {{\omega _j}} \right)} \right],0} \right\} \leqslant \rho $ or $j \leqslant J$

\item \qquad \qquad preset $m = 0$, ${{\mathbf{P}}^{\left( 0 \right)}}\left( {{\omega _j}} \right) = {\mathbf{P}}\left( {{\omega _j}} \right)$, $P_{\text{n}}^{\left( 0 \right)}\left( {{\omega _j}} \right) = {P_{\text{n}}}\left( {{\omega _j}} \right)$;

\item \qquad \qquad While $\left| {{\Upsilon _m}} \right| \leqslant \rho '$ or $m + 1 \leqslant J'$

\item \qquad \qquad \qquad solve (30) with DC programming;

\item \qquad \qquad \qquad solve (31) with DC programming;

\item \qquad \qquad \qquad ${\mathbf{P}}\left( {{\omega _j}} \right) = {{\mathbf{P}}^{\left( {m + 1} \right)}}\left( {{\omega _j}} \right)$;

\item \qquad \qquad \qquad ${P_{\text{n}}}\left( {{\omega _j}} \right) = P_{\text{n}}^{\left( {m + 1} \right)}\left( {{\omega _j}} \right)$;

\item \qquad \qquad \qquad $m = m + 1$;

\item \qquad \qquad End while

\item \qquad \qquad ${\mathbf{P}}_{\left( i \right)}^{{\text{opt}}} = {\mathbf{P}}\left( {{\omega _j}} \right)$;

\item \qquad \qquad $P_{{\text{n,}}\left( i \right)}^{{\text{opt}}} = {P_{\text{n}}}\left( {{\omega _j}} \right)$;

\item \qquad \qquad ${\omega _{j + 1}} = \mu {\omega _j}$;

\item \qquad \qquad $j = j + 1$;

\item \qquad End while

\item \qquad ${\overset{\lower0.5em\hbox{$\smash{\scriptscriptstyle\smile}$}}{\eta } _{\left( {i + 1} \right)}} = \frac{{W{{\overset{\lower0.5em\hbox{$\smash{\scriptscriptstyle\smile}$}}{C} }^{{\text{sec}}}}\left( {{\mathbf{P}}_{\left( i \right)}^{{\text{opt}}},P_{{\text{n,}}\left( i \right)}^{{\text{opt}}}} \right)}}{{{P_{{\text{total}}}}\left( {{\mathbf{P}}_{\left( i \right)}^{{\text{opt}}},P_{{\text{n,}}\left( i \right)}^{{\text{opt}}}} \right)}}$;

\item \qquad $i = i + 1$

\item End while

\item Return ${{\mathbf{P}}^{{\text{opt}}}} = {\mathbf{P}}_{\left( {i - 1} \right)}^{{\text{opt}}}$, $P_{\text{n}}^{{\text{opt}}} = P_{{\text{n,}}\left( {i - 1} \right)}^{{\text{opt}}}$, ${\overset{\lower0.5em\hbox{$\smash{\scriptscriptstyle\smile}$}}{\eta } ^{{\text{opt}}}} = \frac{{W{{\overset{\lower0.5em\hbox{$\smash{\scriptscriptstyle\smile}$}}{C} }^{{\text{sec}}}}\left( {{{\mathbf{P}}^{{\text{opt}}}},P_{\text{n}}^{{\text{opt}}}} \right)}}{{{P_{{\text{total}}}}\left( {{{\mathbf{P}}^{{\text{opt}}}},P_{\text{n}}^{{\text{opt}}}} \right)}}$

                       \end{enumerate}
\STATE \textbf{End Begin}

\end{algorithmic}
\end{algorithm*}

According to the above derivations, secure energy efficiency optimization problem in (21) is firstly transformed into a parametric programming problem (23) based on the Dinkelbach's method. Secondly, the penalty function method is employed to further transform the problem into (26). Next, (26) is solved by alternate searching with DC programming at every step. The derivations for solving the secure EE problem is concluded as the Secure Energy Efficient Hybrid Precoding (SEEHP) algorithm, which is listed at the beginning of this page.

Three layers of iterations are contained in the SEEHP algorithm. Since the convergence of each iteration is specified above, the overall SEEHP terminates with a finite number of iterations. Additionally, based on the hierarchical structure of the SEEHP algorithm, the computational complexity of the SEEHP algorithm is associated with the iteration number in each layer, as well as the computational complexity of the DC programming algorithm used to solve (29) and (30). According to the proofs in \cite{Phan12}, the computational complexity of using the DC programming algorithm to solve (29) and (30) is expressed by

\[\begin{gathered}
  \Delta  = {\Theta _1}O\left( 1 \right)\min \left\{ {\sqrt {\frac{{{\alpha _1}}}{{{\vartheta _1}}}} \ln \left( {\frac{1}{\Omega }} \right),\sqrt {\frac{{{\alpha _1}}}{\Omega }} } \right\} \hfill \\
  \;\;\;\;\;\; + {\Theta _2}O\left( 1 \right)\min \left\{ {\sqrt {\frac{{{\alpha _2}}}{{{\vartheta _2}}}} \ln \left( {\frac{1}{\Omega }} \right),\sqrt {\frac{{{\alpha _2}}}{\Omega }} } \right\} \hfill \\
\end{gathered} ,\tag{32}\]
where $\Omega $ is the DC programming convergence tolerance, ${\Theta _1}$ and ${\Theta _2}$ are the iteration numbers, ${\alpha _1}$ and ${\alpha _2}$ are the Lipschitz constants, ${\vartheta _1}$ and ${\vartheta _2}$ are the convexity parameters for solving (29) and (30), respectively. Further denote the iteration number in each of the three layers as ${i_\upsilon }$, ${j_\rho }$ and ${m_{\rho '}}$, then the overall computational complexity of the proposed SEEHP algorithm is derived as

\[\Xi  = {i_\upsilon }{j_\rho }{m_{\rho '}}\Delta .\tag{33}\]

\begin{table*}[!hbt]
\centering
\caption{\ Simulation parameters of SEEHP algorithm }
\begin{tabular}{l|l}
\hline \textbf{Parameter} & \textbf{Value} \\
\hline
The maximum transmit power ${{P}_{\max }}$ & -5 dBW \\
The minimum secrecy capacity $C_0^{\sec }$ & 3 bit/s/Hz \\
The number of transmit antennas at the gateway controller ${N_{{\text{Tx}}}}$ & 110 \\
The number of receive antennas at the eavesdropper $M$ & 30 \\
The number of RF chains ${N_{{\text{RF}}}}$ & 50 \\
The number of actuators $K$ & 40\\
Static power consumed at the gateway controller ${P_{\text{C}}}$ & 15 W \\
Power consumed by each RF chain ${{P_{{\text{RF}}}}}$ & 30 mW \\
Power consumed by each phase shifter ${{P_{{\text{PS}}}}}$ & -30 dBm \\
Power amplifier efficiency $\alpha $ & 0.38 \\
White complex Gaussian noise power spectral density & -174 dBm/Hz \\
Carrier frequency & 28 GHz \\
Bandwidth $W$ & 20 MHz\\
Path loss exponent $\chi $ & 4.6\\

\hline
\end{tabular}
\label{tab1}

\end{table*}

\section{Numerical results}

In this section, the numerical results of the proposed SEEHP algorithm are illustrated, compared with the secure energy efficient power allocation (SEEPA) algorithm, the secrecy capacity maximization (SCM) algorithm and the hybrid precoding secrecy capacity maximization (HYSCM) algorithm \cite{Zappone16_26,Chen13,Ramadan17}. The SEEPA algorithms aims to maximize the secure energy efficiency with traditional MIMO gateway controller scheme where the number of RF chains equals to the number of transmit antennas \cite{Chen13}. The SCM and HYSCM algorithms maximizes the secrecy capacity with traditional MIMO and hybrid precoding MIMO framework, respectively \cite{Zappone16_26,Ramadan17}. The impacts of the maximum transmit power, the required minimum secrecy capacity, the number of transmit antennas and RF chains at the gateway controller, and the number of the actuators on the secure energy efficiency and secrecy capacity are presented. Without loss of generality, the detailed default simulation parameters are listed in Table I \cite{Chen15_16,HZhang14_24,DWang16_25,Zappone16_26,Zi16}.

\begin{figure}
\vspace{0.1in}
\centerline{\includegraphics[width=9cm,draft=false]{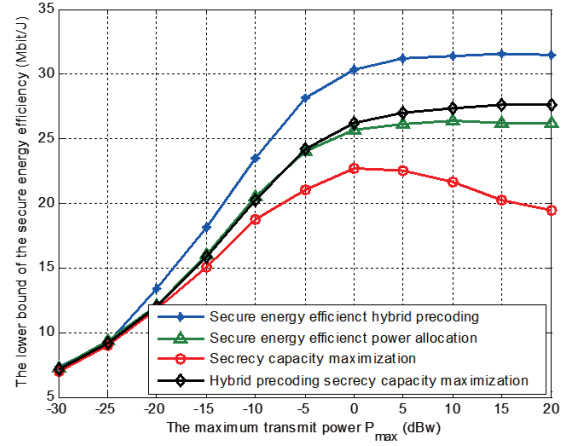}}
\caption{\small \ \ The lower bound of the secure energy efficiency with respect to the maximum transmit power.}
\label{fig2}
\end{figure}

{In Fig. 2, the proposed SEEHP algorithm is compared with the SEEPA algorithm, the SARM algorithm and the SHP algorithm. For each algorithm, numerical results show that the lower bound of the secure energy efficiency firstly increases with the increasing of the maximum transmit power. When the maximum transmit power exceeds a given threshold, the secure energy efficiency of the SEEHP, SEEPA and SHP algorithm eventually saturates. However, the secure energy efficiency of the
SARM algorithm starts to decrease when the maximum transmit power keeps increasing. Compared with the other three algorithms, the proposed SEEHP achieves the highest secure energy efficiency due to the energy efficiency maximization in Section III and the hybrid precoding massive MIMO at the gateway controller. Because more power is consumed by the traditional massive MIMO at the gateway controller and the optimization objective is the secrecy capacity, the SARM algorithm has the lowest energy efficiency. Meanwhile, the SHP and SEEPA algorithms achieve close energy efficiency performance. The simulation results in Fig. 2 reveals that the hybrid precoding scheme significantly improves the secure energy efficiency due to less power consumed at the gateway controller and the relevantly small capacity reduction. Moreover, when the maximum transmit power exceeds a given threshold, the secure energy efficiency stops rising and even starts declining.
}

\begin{figure}
\vspace{0.1in}
\centerline{\includegraphics[width=9cm,draft=false]{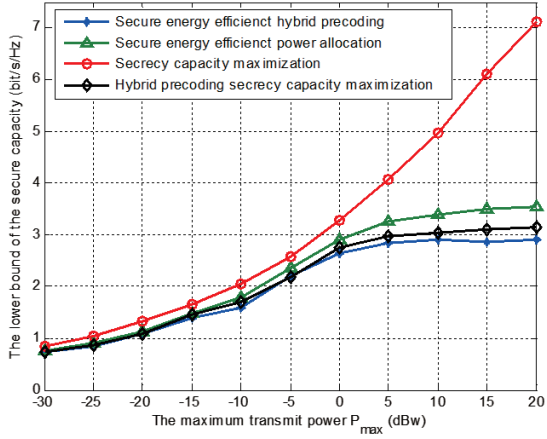}}
\caption{\small \ \ The lower bound of the secrecy capacity with respect to the maximum transmit power.}
\label{fig3}
\end{figure}

Fig. 3 illustrates the lower bound of the secrecy capacity with respect to the maximum transmit power. The SEEPA, SARM and SHP algorithms are also compared with the proposed SEEHP algorithms. According to the numerical results, the lower bound of the secrecy capacity increases with the increasing of the maximum transmit power. Because the optimization objective of the SRAM algorithm is the secure achievable, while the others' are the secure energy efficiency, the SRAM algorithm achieves the highest secrecy capacity performance. The proposed SEEHP algorithm achieves close performance compared with the SEEPA and SHP algorithms. It can be seen in Fig. 3 that the capacity reduction due to the hybrid precoding scheme is quite small, which also explains why the proposed SEEHP algorithm achieves the highest secure energy efficiency in Fig. 2.

\begin{figure}
\vspace{0.1in}
\centerline{\includegraphics[width=9cm,draft=false]{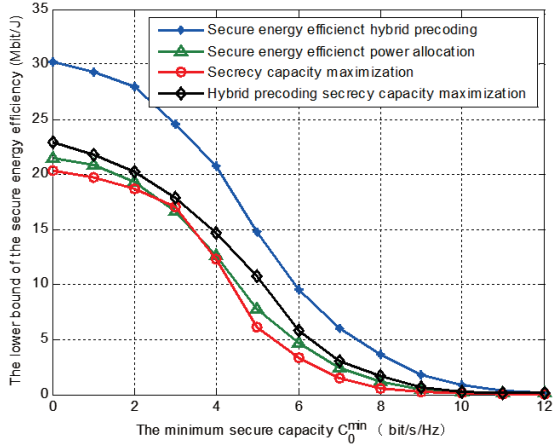}}
\caption{\small \ \ The lower bound of the secure energy efficiency with respect to the minimum secrecy capacity.}
\label{fig4}
\end{figure}

The lower bound of the secure energy efficiency with respect to the minimum secrecy capacity is shown in Fig. 4. When the minimum secrecy capacity increases, the secure energy efficiency performance of the four simulated algorithms deteriorates. This is because the larger minimum secrecy capacity leads to higher outage probability that the achieved secrecy capacity is less than the minimum secrecy capacity. In this case, more transmit power is required to ensure the achieved secrecy capacity is larger than the minimum secrecy capacity, which overall results in lower secure energy efficiency. Compared with the other three algorithms, the proposed SEEHP algorithm achieves the highest secure energy efficiency. It can be seen in Fig. 3 that the capacity reduction due to the hybrid precoding scheme is quite small, which also explains why the proposed SEEHP algorithm achieves the highest secure energy efficiency in Fig. 2.

\begin{figure}
\vspace{0.1in}
\centerline{\includegraphics[width=9cm,draft=false]{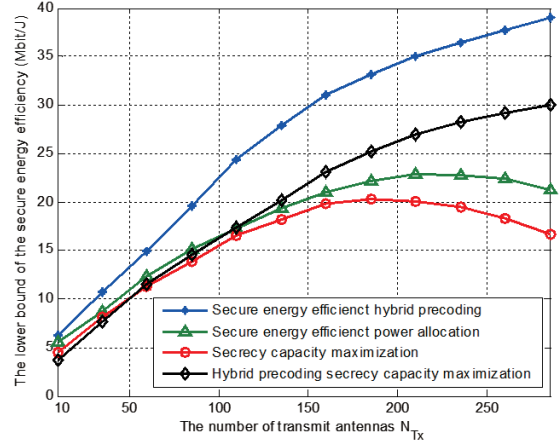}}
\caption{\small \ \ The lower bound of the secure energy efficiency with respect to the number of transmit antennas.}
\label{fig5}
\end{figure}

In Fig. 5, the secure energy efficiency performance is shown with respect to the number of transmit antennas at the gateway controller. The proposed SEEHP algorithm achieves the highest secure energy efficiency, which also rises with the increasing of the number of antennas. Meanwhile, when the number of the antenna is small, the HPSCM algorithm has the lowest energy efficiency. When the number of the antenna is large, both the SEEHP and HPSCM algorithms with hybrid precoding outperform the other two algorithms with traditional MIMO. These results reveal that the increasing of the number of antennas has little impact on the transmit power of the SEEHP and HPSCM algorithms. Thus, exploiting the extra degrees of freedom produced by more antennas and with the transmit power barely increasing, the SEEHP and HPSCM algorithms achieve continuously increasing secure energy efficiency.

\begin{figure}
\vspace{0.1in}
\centerline{\includegraphics[width=9cm,draft=false]{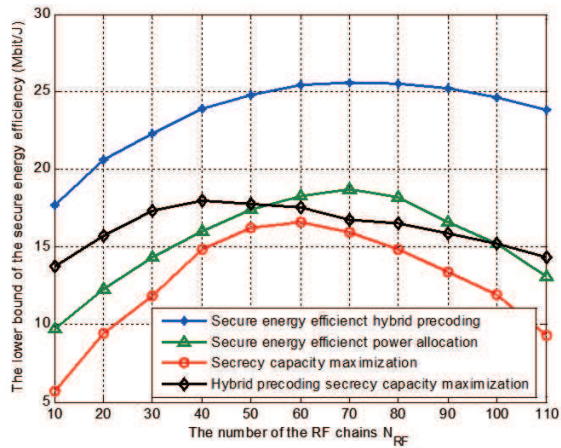}}
\caption{\small \ \ The lower bound of the secure energy efficiency with respect to the number of RF chains.}
\label{fig6}
\end{figure}

The lower bound of the secure energy efficiency with respect to the number of RF chains is shown in Fig. 6. The secure energy efficiency firstly rises then declines with the increasing of the number of RF chains. This fact indicates that when the number of RF chains is small, increasing the number of RF chains contributes to improving the secure energy efficiency. But when the number of RF chains is larger than a threshold, the tremendous power consumed by the RF chains leads to the decreasing of the secure energy efficiency. Comparing with other algorithms, the proposed SEEHP algorithm achieves the highest secure energy efficiency performance. As an algorithm with hybrid precoding scheme, the HPSCM achieves higher performance than the SCM and SEEPA algorithms, especially when the number of RF chains is smaller than 50 and larger than 100.

\begin{figure}
\vspace{0.1in}
\centerline{\includegraphics[width=9cm,draft=false]{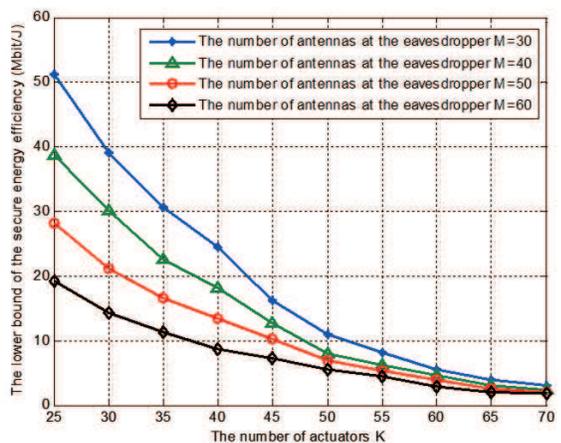}}
\caption{\small \ \ The lower bound of the secure energy efficiency with respect to the number of actuators.}
\label{fig7}
\end{figure}

In Fig. 7, the secure energy efficiency performance of the proposed SEEHP algorithm is illustrated with respect to the number of antennas at the eavesdropper. The secure energy efficiency decreases with the increasing of the number of actuators. When the number of actuators is larger than 50, which is the number of RF chains, the secure energy efficiency becomes quite small and gradually approaches 0. Meanwhile, the larger number of antennas at the eavesdropper corresponds to lower secure energy efficiency. This is due to that increasing the number of antennas at the eavesdropper leads to the increasing of the received SINR at the eavesdropper in (5), which leads to the decreasing of the secrecy capacity and the secure energy efficiency.

The above results from Fig.2 to Fig. 7 together reveal the fact that for the IoT network with a massive MIMO gateway controller, multiple actuators and an eavesdropper, the secure energy efficiency can be significantly improved by implementing hybrid precoding scheme and secure energy efficiency optimization. Moreover, with hybrid precoding scheme, increasing the number of transmit antennas leads to little impact on the transmit power. Then the secure energy efficiency benefits from the increasing of the degrees of freedom produced by more antennas with the transmit power barely increases.

\section{Conclusions}
In this paper, the SEEHP algorithm is proposed to improve the secure energy efficiency of the IoT network gateway controller with hybrid precoding massive MIMO. Considering the power consumption by the transmitted signals and active noise as well as the gateway controller hardware, the secure energy efficiency optimization problem is formulated. Due to the non-convexity of the problem and the feasible domain, RF and baseband precoding schemes are then designed. The problem then transforms into a suboptimal form with transmit power of the signals and active noise to be optimized. To further solve the transformed problem, the Dinkelbach's method, the penalty function method and the DC programming method are utilized in a hierarchical structure, which in general forms the SEEHP algorithm. Numerical results shows that the proposed SEEHP algorithm achieves the highest secure energy efficiency compared with other three physical layer security algorithms. When the maximum transmit power is small, the SEEHP algorithm achieves close secrecy capacity compared with the other three algorithms. In addition, when the number of antennas keeps increasing, the SEEHP algorithm achieves much higher secure energy efficiency than the compared algorithms. In the near future, the authors will try to investigate and optimize the secure energy and spectral efficiency of the IoT network with multiple eavesdroppers and imperfect CSI.

\end{document}